\documentclass[11pt,twoside]{article}
\usepackage{asp2014}

\aspSuppressVolSlug
\resetcounters

\begin{document}

\title{CEFCA Catalogues Portal towards FAIR principles}

\author{Tamara~Civera$^1$}
\affil{$^1$Centro de Estudios de F{\'i}sica del Cosmos de Arag{\'o}n (CEFCA), Teruel, Arag{\'o}n, Spain; \email{tcivera@cefca.es}}

\paperauthor{Tamara~Civera}{tcivera@cefca.es}{}{Centro de Estudios de F{\'i}sica del Cosmos de Arag{\'o}n (CEFCA)}{UPAD}{Teruel}{Arag{\'o}n}{44001}{Spain}



  
\begin{abstract}

The Centro de Estudios de F{\'i}sica del Cosmos de Arag{\'o}n (CEFCA) is carrying out from the Observatorio Astrof{\'i}sico de Javalambre (OAJ, Teruel, Spain) two large area multiband photometric sky surveys, J-PLUS and J-PAS, covering the entire optical spectrum using narrow and broad band filters. J-PAS and J-PLUS include coadded and individual frame images and dual and single catalogue data. To publish all of this data, the CEFCA catalogues portal has been implemented offering web user interface services, as well, as Virtual Observatory (VO) services.

This contribution presents the effort and work done in the CEFCA Catalogues Portal to enhance data publication of these large surveys following FAIR principles to increase data value and maximize research efficiency. It presents how FAIR principles have been achieved and improved with the implementation and publishing of the CEFCA Catalogues Publishing Registry, the use of VO services, their validation and improving processes and the effort made to offer data to improve provenance information.
  
\end{abstract}


\section{The Observatorio Astrof{\'\i}sico de Javalambre (OAJ)}

The Centro de Estudios de F{\'\i}sica del Cosmos de Arag{\'o}n (CEFCA\footnote{\url{http://www.cefca.es}}) is a center for research in Astrophysics and Cosmology whose activities focus on the technological development and operation of the Observatorio Astrof{\'i}sico de Javalambre (OAJ\footnote{\url{http://oaj.cefca.es/}}) and on its scientific exploitation. 

The OAJ has been particularly conceived for carrying out large sky surveys with two large field telescopes: JST250, a 2.5m 3deg FoV and JAST80 a 80cm 2deg FoV. The most immediate objective of the two telescopes for the next years is carrying out two unique multiband photometric sky surveys of 8500 square degrees: Javalambre Physics of the Accelerating Universe Astrophysical Survey (J-PAS; \citet{2019AAS...23338301D}; \citet{bonoli:hal-03129632}) (with JST250 using 54 narrow plus 5 broad band filters) and Javalambre Photometric Local Universe Survey (J-PLUS; \citet{2019A&A...622A.176C}) (with JAST80 using 12 filters). The data archiving, processing  and publication of all of these images and its products is carried out by the Unit for Processing and Data Archiving (UPAD; \citet{2014SPIE.9152E..0OC}).

\section{The CEFCA Catalogues Portal}

J-PAS and J-PLUS include images, dual and single catalogue data which include parameters measured from images and photo-redshift computations. A powerful web portal, the CEFCA Catalogues Portal (archive.cefca.es; (\citet{P2-12_adassxxix})), has been implemented to publish all this survey data offering advanced tools, each suited to a particular need, for data search, visualization and download. This portal includes web user interface services such as sky navigator, object visualization, object list search, ADQL asynchronous queries interface, cone search and image search and download. Offering also all of this data through Virtual Observatory (VO) services like SIAP (Simple Image Access Protocol), SCS (Simple Cone Search), TAP (Table Access Protocol) and catalogue and images HIPS (Hierarchical Progressive Survey).


\section{Towards FAIR principles}

These large projects require an easy and agile access to data, data interoperability and reusability as well as ensure its discovery (FAIR principles \footnote{\url{https://www.go-fair.org}}) to give data greater value, enhance their propensity for reuse, by humans and at scale by machines and maximize research efficiency (\citet{20.500.12259_103794}). For this, a great effort and work has been done on the CEFCA Catalogues Portal to enhance surveys data publication following FAIR principles and making it findable, accessible, interoperable and reusable.

\subsection{Findable}

Data is findable when it is described by rich metadata and it is registered or indexed in a searchable resource that is known and accessible to potential users. Additionally, a unique and persistent identifier should be  assigned  such  that  the  data  can  be  unequivocally  referenced  and  cited  in  research  communications. 

To make OAJ surveys data findable, a VO harvest registry, the CEFCA Catalogues Publishing Registry has been implemented registering, indexing and offering each survey data release as a different VO Resource \footnote{\url{https://www.ivoa.net/documents/VOResource/index.html}} (F4 principle). In each VOResource, the survey data release data is rich and in detail described as well as all the services to access it (F2 principle).

To globally, unique and persistent identify each survey data release an IVOA identifier (IVOID) has been assigned to each of them (F1 principle). This IVOID is specified in each VOResource and used by the CEFCA Catalogues Publishing Registry to index them (F3 principle).


\begin{table}[!ht]
\caption{IVOIDs of the current data releases published}
\smallskip
\begin{center}
{\small
\begin{tabular}{llc}  
\tableline
\noalign{\smallskip}
J-PLUS DR2 & ivo://CEFCA/j-plus/J-PLUS-DR2 \\
J-PLUS DR1 & ivo://CEFCA/j-plus/J-PLUS-DR1 \\
MiniJ-PAS PDR201912 & ivo://CEFCA/minijpas/MINIJ-PAS-PDR201912 \\
\noalign{\smallskip}
\tableline\
\end{tabular}
}
\end{center}
\end{table}

Furthermore, the CEFCA Catalogues Publishing Registry has been registered in the IVOA Registry of Registries (RofR) \footnote{\url{http://rofr.ivoa.net/}} what allows VO applications like Aladin \footnote{\url{https://aladin.u-strasbg.fr/}} or TOPCAT \footnote{\url{http://www.star.bris.ac.uk/~mbt/topcat/}} to find and directly use OAJ surveys data and services.

\articlefiguretwo{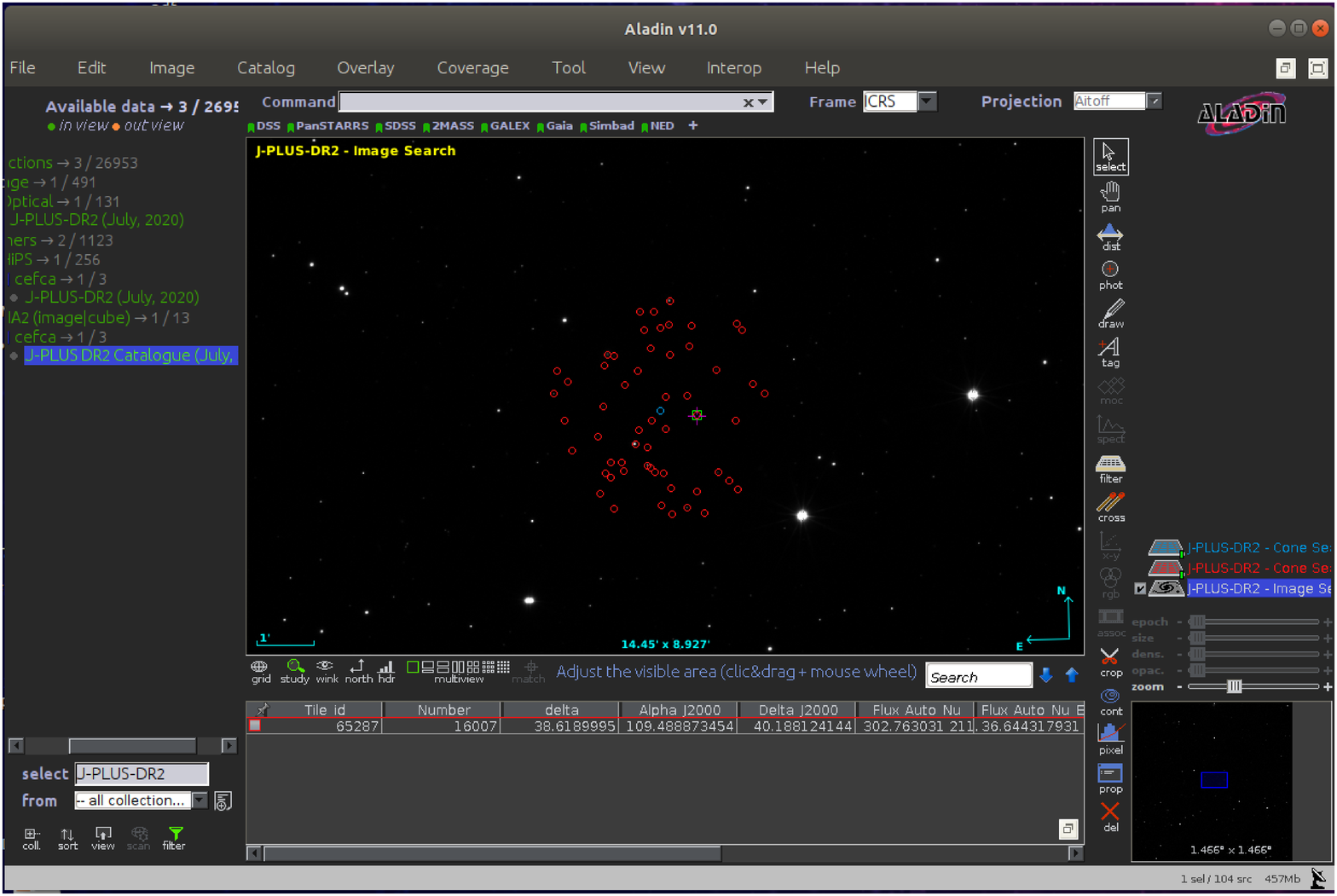}{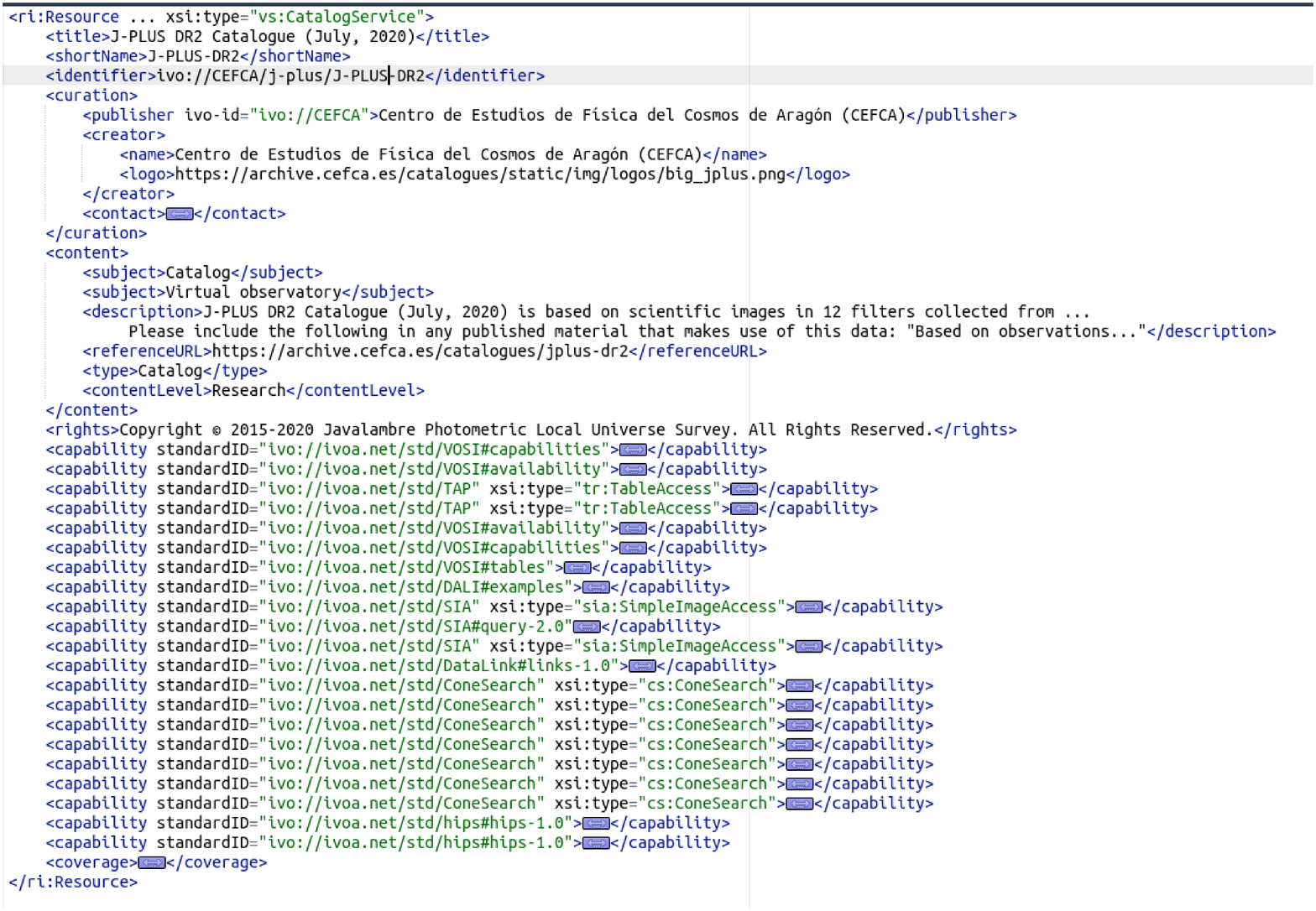}{fig1}{\emph{Left:} Aladin application using data of J-PLUS DR2. \emph{Right:} J-PLUS DR2 VOResource.}

\subsection{Accessible}

Data is accessible when it can be obtained by humans and machines through a well-defined, open, free, and universally implementable protocol.

All surveys data (images and catalogues data) is accessible through different VO services like SIAP, SCS, TAP and HIPS, so it is offered using open, free and universally implementable protocols (A1 principle). Furthermore, to improve data access all the CEFCA Catalogue VO services implementations have been checked through different external IVOA validators and improved to be full compliant with the protocols (\citet{IVOAMay2021}). 

All surveys data releases metadata is always accessible through CEFCA Catalogues Publishing Registry even when data is no longer available (A2 principle).

\subsection{Interoperable}

Data usually need to be integrated with other data. In addition, data need to interoperate with applications for analysis, storage, and processing. Interoperable data and metadata must use a formal, accessible, shared, and broadly applicable language for knowledge representation. They use vocabularies which themselves follow the FAIR principles, and they include qualified references to other data or metadata.

In the Catalogues Portal, data and metadata are returned in VO formats so a formal, accessible, shared and broadly applicable language for knowledge representation is used. They also use VO vocabularies and semantics like Universal Content Descriptors (UCDs) and UTypes so vocabularies that follow FAIR principles are used (I1 principle). This data and metadata also have been checked through different external IVOA validators and improved to be full compliant with the protocols. In addition, all the surveys data can be downloaded in standard and accessible formats like CSV, FITS and VOTable.
    		       					 
Data and metadata include qualified references to other data. For example, in the CEFCA Catalogues registry all the data releases of the same survey are interconnected and users can search by survey (set) (I2 principle).

Furthermore, all the web interface services support Simple Application Messaging Protocol (SAMP) that
enables the CEFCA catalogues portal interoperate and communicate with VO-compatible applications.

\subsection{Reusable}

For data to be reusable, the FAIR principles reassert the need for rich metadata and documentation that meet relevant community standards and provide information about provenance. This covers reporting how data was created and information about data reduction or transformation processes to make data more usable, understandable or 'science-ready'.

Each survey data release data is richly described in the VOResource and all the metadata returned in the VO services results has been improved to offer as many attributes as possible (R1 principle). In the VOResource data, the acknowledge to include in any published material that makes use of the data is indicated. In addition, to improve the provenance information, the metadata and data of the individual frame images used to generate the coadded images are also provided through different VO services like SIAP, SCS and TAP.

\section{Conclusions}

The work done in the CEFCA Catalogues Portal to enhance data publication of the OAJ surveys data following FAIR principles to increase data value and maximize research efficiency has been presented. As well as how the 
implementation of the CEFCA Catalogues Publishing Registry, the use of VO services, the validation and improving processes of them and the publication of the individual frame images data have improved them. But there is still work to do to make it more FAIR. This future work includes assigning DOIs to each survey data release, studying which data use license is more suitable or offering more provenance information.

\acknowledgements Funding: Fondo de Inversiones de Teruel, PGC2018-097585-B-C21 (MCIU/AEI/FEDER, UE)

\bibliography{X4-001}


\end{document}